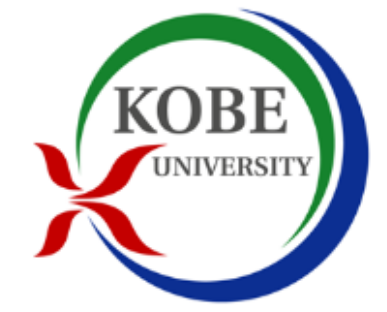

# Advancing the Physical Internet with GraphRAG: A New Way to Review and Integrate Existing Research

Hisatoshi Naganawa[1], Enna Hirata[2], Russell G. Thompson[3], and Akira Yamada[4]

1. Faculty of Ocean Science and Technology, Kobe University, Kobe 658-0022, Japan
2. Graduate School of Maritime Sciences, Kobe University, Kobe 658-0022, Japan
3. Department of Infrastructure Engineering, the University of Melbourne, Australia.
4. Center for Mathematical and Data Sciences, Kobe University, Kobe 657-8501, Japan.

Corresponding author: enna.hirata@platinum.kobe-u.ac.jp

***Abstract:*** *Physical Internet (PI) is an emerging concept that applies the digital internet as a design metaphor for the development of sustainable, interoperable, and collaborative freight transportation. It is considered a way to bring logistics into the next generation of transformation. Effective tools for organizing and integrating knowledge are essential for navigating the emerging research in this area. In this study, we explore the application of Graph Retrieval Augmented Generation (GraphRAG) in the context of PI, using GPT-4o mini and Neo4j to construct a knowledge graph and systematically analyze existing PI-related literature. Our approach synthesizes scattered research findings, highlights emerging trends, and identifies knowledge gaps. Furthermore, we demonstrate that GraphRAG improves accessibility by structuring complex information into interconnected graphs and provides a deeper understanding of underlying research dynamics. This research will contribute to future research and innovation by providing a new method of information analysis in the PI domain.*

## 1 Introduction

Physical Internet (PI) refers to the concept of creating a sustainable global logistics network by incorporating the concept of the digital internet into the transportation and distribution of physical goods. It aims to improve efficiency, flexibility, and sustainability by redesigning existing logistics networks and utilizing information and communication technologies (ICT) alongside standardized logistics infrastructures. Almost fifteen years have passed since Montreuil et al. (2010) first introduced the PI concept, which has since progressed from a theoretical framework to active empirical research. Efforts to make PI a reality have gained momentum through international conferences (IPIC) and the development of roadmaps by national governments, including EU and Japan.

In this study, we apply a graph retrieval-augmented generation (GraphRAG) approach to gain deeper insights from the existing literature. GraphRAG is an innovative method that combines information retrieval and generation technologies. It helps clarify relationships between data, efficiently retrieves the most relevant information in response to a query, and then generates new contents based on these results. This approach delivers powerful performance, especially for solving problems with complex data structures.

We aim to assess the potential contributions of existing structured research and theoretical frameworks by leveraging GraphRAG to analyze them. Specifically, we demonstrate how GraphRAG can support policy development and the formulation of research agendas to realize PI. This is the first application of GraphRAG in the context of PI-related studies. It offers a novel perspective and contributes to advancing this research domain.

The remainder of this paper is organized as follows: Section 2 presents a literature review. Section 3 describes the data used and the proposed methodology. Section 4 presents several results and compares them. Section 5 concludes and identifies areas of research.

# 2 Literature Review

## 2.1 Brief overview on Physical Internet

The concept of PI was introduced by Montreuil in 2010 (Montreuil et al., 2010). PI is an innovative concept that uses the digital internet as a design metaphor to foster the development of sustainable, interoperable and coordinated freight transportation (Sternberg and Norrman, 2017). The key concept behind the PI logistics system is to route highly modular containers to transit centers, known as PI hubs, to achieve a highly efficient transportation network that takes advantage of consolidation opportunities (Venkatadri et al., 2016). The concept of PI has attracted many stakeholders to support the development of logistics networks in the last few years; the literature on PI has increased dramatically over the past decade. Although summaries of past and future PI research are sporadically reported in recent studies (Aron and Sgarbossa, 2023; Ballot et al., 2021), now fifteen years after the introduction of PI, there is a growing need for an integrated method to efficiently organize the current research and address future research challenges.

## 2.2 Brief overview on generative AI

Artificial intelligence (AI) has attracted great attention in a variety of fields and industries (Hyder et al., 2019). Beginning with Alan Turing's basic concepts in the 1950s, AI has long gone through several stages of development and a period of stagnation, known as the AI winters (Turing, 2009). However, the global prominence of AI surged after OpenAI released its chat generation pre-trained transformer (ChatGPT) in late 2022. The GPT family uses huge data sets and large language models (LLMs) trained with deep learning techniques to generate human-like text (Cascella et al., 2024). GPTs excel at understanding context and processing diverse queries, but their main limitation is knowledge rigidity. It relies on existing training data and cannot incorporate real-time updates. Furthermore, when faced with unknown information, they struggle to provide factually accurate responses, often misleading or deriving incorrect content through inference. These limitations pose challenges for applications that require high reliability (McIntosh et al., 2024; Oviedo-Trespalacios et al., 2023; Rawte et al., 2023).

## 2.3 GraphRAG

One potential solution to this challenge is the use of GraphRAG, which has attracted considerable interest. GraphRAG is a technology that improves the accuracy of responses by adding the ability to search external databases to a LLM (Figure 1). Developed by Microsoft researchers, GraphRAG was introduced in February 2024, and a reference implementation was made available on Github in July 2024 (Microsoft, 2025).

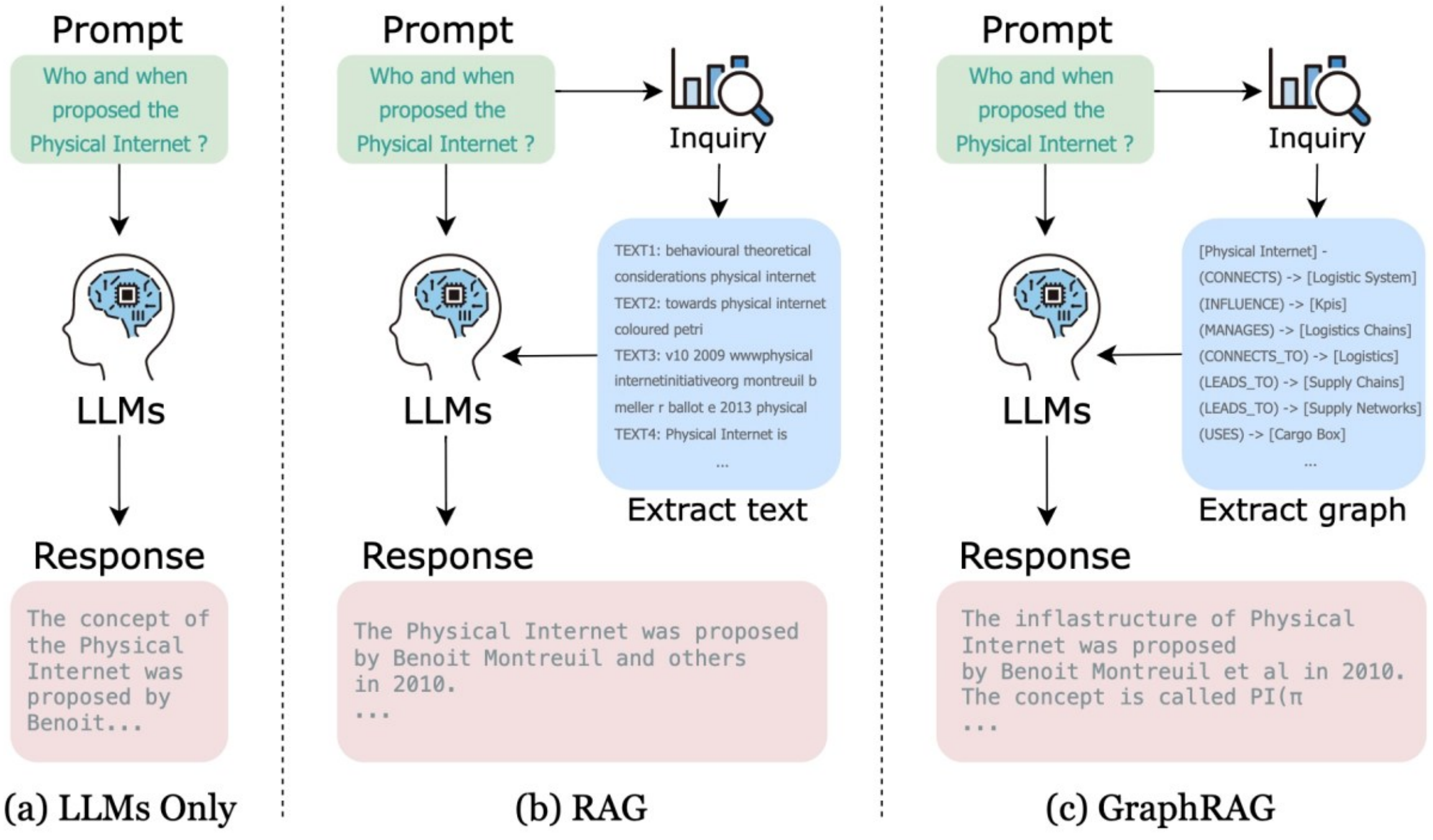

*Figure 1: Different methods using LLMs*

GraphRAG is an advanced form of retrieval-augmented generation (RAG) that enhances large language models by structuring retrieved information as a graph. In this graph, nodes represent concepts, and edges denote their relationships. This graph-based approach allows the model to better understand context, identify meaningful connections, and reason more effectively across multiple documents. It is particularly useful for complex tasks such as research synthesis and policy development. GraphRAG is relevantly new technology with a limited number of studies exploring its application. Li et al. (2025) propose to use GraphRAG to convert vast amounts of unstructured supplier capability data into a knowledge graph, thereby improving supplier discovery and making manufacturing suppliers more accessible and searchable. Ojima et al. (2024) present a knowledge management system for automotive failure analysis using GraphRAG.

## 3 Data

In this study, we create a knowledge graph (KG) specific to PI by integrating three data sets (Table 1). The first set consists of the proceedings of the international physical internet conference (IPIC) held from 2016 to 2024. The second set is the PI roadmap developed by ALICE (ALICE, 2020), and the third set is the PI roadmap for Japan prepared by the Ministry of Economy, Trade and Industry (METI, 2022). The IPIC proceedings were downloaded from the conference website since its inception; the ALICE PI roadmap was obtained from the ALICE website; and the Japan PI roadmap was obtained from the METI website. A Python script is used to convert PDF documents into text data for subsequent analysis.

*Table 1: List of Contents*

| Contents | Year | Source and collection method |
|---|---|---|
| Proceedings of IPIC | 2016 ~ 2024 | Download from IPIC repository (IPIC, 2025) |
| PI Roadmap (ALICE) | 2020 | Download from ALICE website (ALICE, 2020) |
| PI Roadmap (Japan) | 2022 | Download from METI website (METI, 2022) |

# 4 Methodology

The overall research flow is shown in Figure 2. It consists of data collection, preprocessing, response generation, and performance measurement.

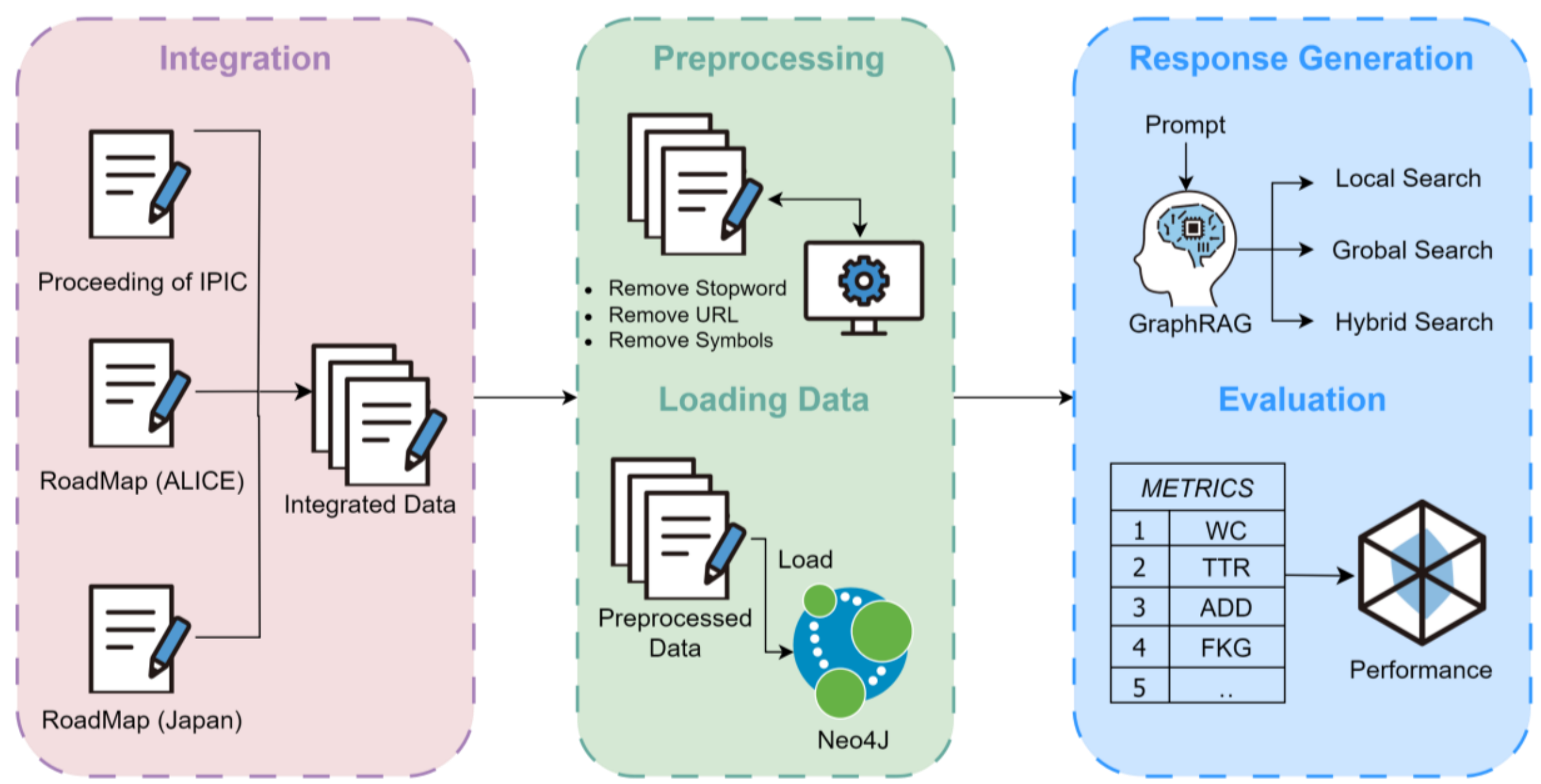

*Figure 2: Different methods using LLM*

## 4.1 Preprocessing

For preprocessing, we first integrate IPIC proceedings data and data from ALICE and Japan's PI roadmaps. The NLTK library is then applied to remove stop words, which include standard stop words and content specific stop words such as URLs, symbols, and other extraneous characters (*NLTK*). The resulting refined text data is then loaded into GraphRAG.

## 4.2 Response generation based on GraphRAG

GraphRAG is used to generate responses to the two prompts in four models: GPT-Only (I), Local Search (II), Global Search (III), and Hybrid Search (IV). Local Search (LS) model searches for nodes and edges in the graph that are semantically close to the query and retrieves directly related items from the graph database. Global Search (GS) model identifies clusters within the graph and provides a summarized view of the entire community relevant to the query. Hybrid Search (HS) model combines LS and GS to generate comprehensive responses with both specific examples and broader contextual information.

GraphRAG was created using neo4j desktop as the database and GPT 4o-mini as the LLM; the two prompts were created from two perspectives: promoting innovation and research, and practical implementation and social impact (Table 2).

*Table 2: Prompt*

| No. | Prompt |
|---|---|
| 1 | Comparing to the ALICE and Japan roadmaps, what are the areas lacking in research focusing on Physical Internet? |
| 2 | Please provide concrete proposals for promoting the social implementation of the Physical Internet in Japan. |

## 4.3 Performance metrics

The performance of each model is evaluated using five metrics (Table 3). Paraphrase-MiniLM-L6-v2 was applied to embed sentences, for the calculation of cosine similarity.

*Table 3: Performance Evaluation Metrics*

| Metrics | Abbreviation | Description |
|---|---|---|
| Word Count | WC | Total number of words in a sentence. |
| Type Token Ratios | TTR | Measures vocabulary diversity. |
| Average Dependency Distance | ADD | Measures the distance of dependencies between words in a sentence. |
| Flesch-Kincaid Grade Level | FKG | Indicator for evaluating the readability. |
| Cosine Similarity | CS | An index that quantifies the degree of textual similarity. In this study, it is calculated based on responses generated by Model I (GPT-only). |

# 5 Results and Discussion

## 5.1 Prompts and Reponses

Following the definition explained in Section 4.2, the prompts are run in each of the four models, namely (I) GPT-Only, (II) Local Search, (III) Global Search, and (IV) Hybrid Search, and the results are shown in Table 4, 5.

*Table 4: Response to prompt 1- Comparing to the ALICE and Japan roadmaps, what are the areas lacking in research focusing on physical internet?*

| Model | Output |
|---|---|
| I | While the ALICE and Japan roadmaps lay a solid foundation for the Physical Internet (PI), several research gaps remain. Key areas include the development of universal standards for interoperability, secure data-sharing mechanisms, and more robust sustainability metrics. There is also limited research on decentralized logistics models, which could improve resilience. Emerging technologies like IoT, AI, and blockchain need further exploration for integration into the PI framework. Additionally, workforce impacts, including job displacement and retraining, are insufficiently addressed. Lastly, economic models and funding mechanisms for PI adoption require more in-depth study. |
| II | The areas lacking in research focusing on the Physical Internet, when comparing the ALICE and Japan roadmaps, include a more comprehensive exploration of reinforcement learning applications, integration of logistics providers, and strategies for achieving emission targets. Additionally, there is a need for enhanced collaboration between stakeholders and a clearer definition of research innovation paths specific to the Physical Internet. |
| III | The areas lacking in research focusing on the physical internet, compared to the ALICE and Japan roadmaps, include insufficient exploration of protocols and methods for implementing digital internet functionalities within the physical internet framework. Additionally, there is a need for deeper analysis of resilience and efficiency in the transport of physical goods, as well as a focus on integrating IoT technologies for improved risk management in construction and logistics. Overall, there is a gap in addressing the complexities of logistics networks and the development of commercial data platforms that ensure seamless and transparent operations. |
| IV | The areas lacking in research focusing on the physical internet, compared to the ALICE and Japan roadmaps, include insufficient exploration of container development, hub inventory management, and the integration of Internet of Things (IoT) technologies in risk management within prefabricated construction. Additionally, there is a need for deeper analysis of protocols and methods for resilience and efficiency in transport concepts, as well as a more comprehensive understanding of the functionalities and attributes of road-based physical internet systems. |

*Table 5: Response to prompt 2 - please provide concrete proposals for promoting the social implementation of the Physical Internet in Japan.*

| Model | Output |
|---|---|
| I | To promote the social implementation of the Physical Internet in Japan, it is essential to foster collaboration between government, industry, and academia to create a unified vision. Developing pilot projects in urban areas can demonstrate the benefits of interconnected logistics systems. Encouraging |
| | public-private partnerships will facilitate investment in necessary infrastructure and technology. Raising awareness through educational campaigns can help inform stakeholders about the advantages of the Physical Internet. Additionally, establishing regulatory frameworks that support innovation while ensuring safety and security is crucial. Finally, leveraging Japan's advanced technology landscape, such as IoT and AI, can enhance the efficiency and effectiveness of the Physical Internet. |
| II | To promote the social implementation of the Physical Internet in Japan, it is essential to foster collaboration among stakeholders, including government, industry, and academia. Encouraging public awareness campaigns can help educate citizens about the benefits of the Physical Internet. Additionally, investing in infrastructure that supports seamless logistics and data sharing will be crucial. Implementing pilot projects can demonstrate the effectiveness of the Physical Internet in real-world scenarios, while providing incentives for businesses to adopt these practices can drive participation. Finally, establishing regulatory frameworks that facilitate innovation and ensure data security will support sustainable growth in this area. |
| III | To promote the social implementation of the Physical Internet in Japan, it is essential to enhance coordination and collaboration among stakeholders to build social capital. Encouraging transparency and clear communication about the tangible benefits of the Physical Internet will help engage businesses and researchers alike. Highlighting successful case studies can demonstrate the effectiveness and value of the Physical Internet, fostering trust and cooperation. Additionally, addressing potential barriers through education and awareness initiatives will be crucial in attracting new participants to the network. Finally, focusing on the financial aspects and quality outcomes will ensure that stakeholders recognize the importance of collaboration for future success. |
| IV | To promote the social implementation of the Physical Internet in Japan, it is essential to enhance stakeholder engagement by showcasing successful case studies that demonstrate tangible benefits and quality outcomes. Encouraging collaboration among researchers, businesses, and logistics providers can foster a shared understanding of the value of the Physical Internet. Emphasizing the importance of clear communication and education about costs and operational decisions will help address potential barriers. Additionally, building trust and reciprocity within logistical networks can strengthen social capital, facilitating the transition to open supply networks. Finally, leveraging innovative technologies, such as reinforcement learning for logistics optimization, can improve efficiency and sustainability in urban logistics. |

The intuitive impression is that the GPT model (I) generate more broader, general proposals, while RAG-based models (II, III, IV) provide more specific and targeted suggestions. Quantitative performance evaluation is discussed in Section 5.2.

## 5.2 Performance Metrics

The performance of each model, evaluated using the metrics described in Section 4.3, is shown in Table 6. The results show that the GraphRAG models (Models II – IV) outperformed the standard GPT model (Model I) on all metrics except WC. In particular, Model IV (HS) achieved highest scores of all models. This is likely because the standard GPT model is superior at understanding broader, context-rich scenarios and generating responses in relatively plain language, whereas GraphRAG incorporates specialized terminology and complex vocabulary from external knowledge sources, resulting in more detailed and domain-specific responses. In other words, GraphRAG can be particularly valuable when factual accuracy and transparent source traceability are critical. To advance research and real-world implementation of PI, the optimal model should combine a search-based system like GraphRAG for factual grounding with the GPT's ability to synthesize information and foster creative system design.

*Table 6: Performance Metrics (bold indicates best performance)*

| **Prompt** | **1** | | | | **2** | | | |
|---|---|---|---|---|---|---|---|---|
| **Model** | **I** | **II** | **III** | **IV** | **I** | **II** | **III** | **IV** |
| WC | **108** | 66 | 102 | 87 | 117 | 109 | 112 | **119** |
| TTR | 0.72 | **0.73** | 0.64 | 0.68 | 0.68 | 0.70 | 0.67 | **0.71** |
| ADD | 3.19 | 2.94 | 3.16 | **3.41** | 2.99 | **3.09** | 2.99 | 2.91 |
| FKG | 15.00 | 21.90 | 20.20 | **24.20** | 18.30 | 19.10 | 18.40 | **19.80** |
| CS | - | 0.67 | **0.71** | 0.67 | - | **0.95** | 0.86 | 0.87 |

## 6 Conclusion

This study provides a new perspective on the research challenges and social implementation of PI by analyzing large amounts of textual data related to the PI using GraphRAG. Going beyond traditional statistical analysis, this study provides a new approach to understanding the challenges of implementing the PI. Four models were used to generate responses to two prompts representing the innovation and research promotion, and the practical application and social impact perspectives. In both cases, GraphRAG outperformed the GPT model, demonstrating higher lexical density, readability, and similarity of responses.

This study makes several important contributions. First, it presents the first application of GraphRAG to PI, pioneering its use in reviewing and synthesizing existing research. The results demonstrate the potential of AI-driven methods in providing new insights from existing literature. Second, by applying a large-scale natural language processing approach, this study provides a solution for capturing diverse datasets, including not only academic literature but also real time internet data sources reflecting public opinion and industry perspectives. Third, by comparing various search strategies (local, global, and hybrid) within GraphRAG, this study provides valuable insights into the optimization of search augmentation generation models for PI.

There are several limitations to this study. First, the dataset consists primarily of data from IPIC proceedings. While the proceedings provide relevant insights into PI, they lack the technical depth and detailed information associated with the topic. Second, the use of GraphRAG, especially hybrid search, increases computational costs, potentially limiting its scalability for real-time search augmentation. Third, the study does not compare its results to official logistics data or expert opinion, which limits the validation of its conclusions.

Future research should combine GraphRAG with broader range of academic papers, official industry reports, and other sources to generate more comprehensive insights. In addition, the development of improved methods for validating and interpreting AI-generated insights would be valuable in advancing the adoption and implementation of PI. Finally, further research should explore how AI-driven analytics can influence government regulations and industry standards for sustainable logistics.

## 7 Conflicts of Interest

The authors declare that the research was conducted in the absence of any commercial or financial relationships that could be construed as a potential conflict of interest.

## 8 Acknowledgements

This research supported by JSPS KAKENHI Grant Numbers JP23K04076.

Appendix

Figure A1 shows a visualization of the knowledge graph in GraphRAG on Neo4j Desktop. The number of nodes is limited to 100. Nodes represent data entities, such as people or products, and edges represent relationships between nodes, such as friendships or purchasing relationships. Nodes are automatically color-coded according to their type.

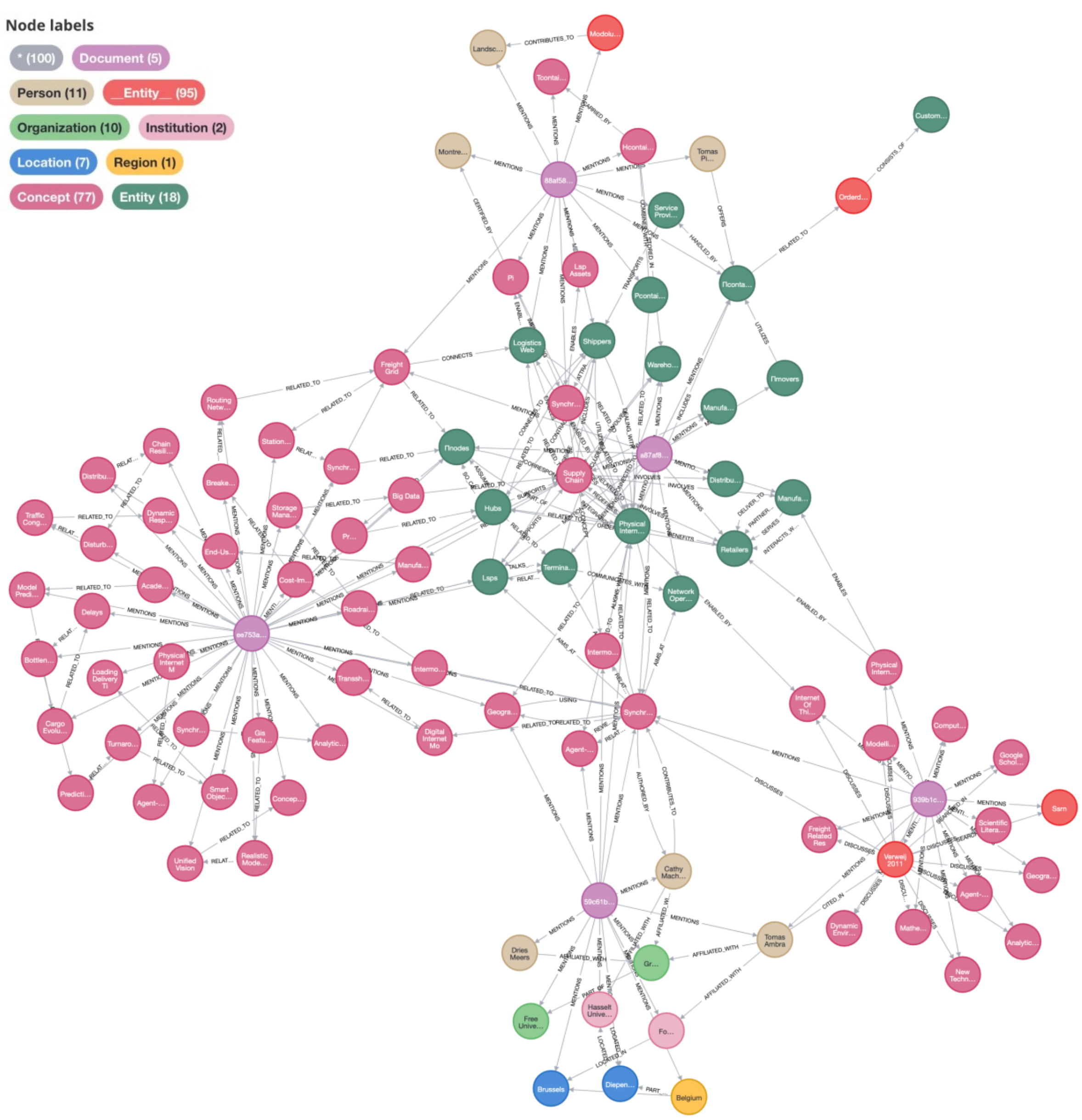

*Figure A1: Knowledge Graph (Limit 100 nodes)*